\documentclass[12pt]{article}

\usepackage{amssymb}
\usepackage{amsmath}
\usepackage{amscd}
\usepackage{latexsym}
\usepackage{graphicx}

\newcommand{\be}{\begin{equation}}
\newcommand{\ee}{\end{equation}}

\newcommand{\vp}{\varphi}

\newcommand{\ra}{\rightarrow}

\newcommand{\cL}{{\cal L}}
\newcommand{\rgl}{\rangle}
\newcommand{\lgl}{\langle}

\begin{document}

\begin{center}

{\Large{\bf Conditions for Quantum Interference\\ in Cognitive Sciences} \\ [5mm]

V.I. Yukalov$^{a,b}$ and D. Sornette$^{a,c}$} \\ [3mm]

{\it
$^a$Department of Management, Technology and Economics, \\
ETH Z\"urich (Swiss Federal Institute of Technology) \\
Z\"urich CH-8092, Switzerland \\
e-mail: syukalov@ethz.ch \\ [3mm]

$^b$Bogolubov Laboratory of Theoretical Physics, \\
Joint Institute for Nuclear Research, Dubna 141980, Russia \\ 
e-mail: yukalov@theor.jinr.ru \\ [3mm]

$^c$Swiss Finance Institute, c/o University of Geneva, \\
40 blvd. Du Pont d'Arve, CH 1211 Geneva 4, Switzerland \\
e-mail: dsornette@ethz.ch}

\end{center}

\vskip 1cm

{\parindent=0pt
{\bf Keywords}: Decisions under uncertainty; Quantum measurements;
Quantum probability; Conditions for interference

\vskip 1 cm

{\bf Corresponding author}: V.I. Yukalov

Department of Management Technology and Economics, ETH Z\"urich

Scheuchzerstrasse 7, Z\"urich CH-8092, Switzerland

\vskip 2mm
{\bf Phone}: 41 (44) 632-9282

\vskip 2mm
{\bf E-mail}: yukalov@theor.jinr.ru

}

\newpage

\begin{abstract}
We present a general classification of the conditions under which cognitive 
science, concerned e.g. with decision making, requires the use of quantum theoretical 
notions. The analysis is done in the frame of the mathematical approach based 
on the theory of quantum measurements. We stress that quantum effects in 
cognition can arise only when decisions are made under uncertainty. 
Conditions for the appearance of quantum interference in cognitive sciences 
and the conditions when interference cannot arise are formulated.
\end{abstract}

\newpage

\section{Introduction}

The special issue, devoted to quantum-like effects in cognition, is certainly,
of great interest giving hope that such formulations could explain a variety 
of empirical phenomena that do not find explanations in classical terms. The 
purpose of the special issue is well described in the introductory article by
Wang, Busemeyer, Atmanspacher, and Photos (2013). All papers in this issue
appeal, in one way or another, to the existence of quantum interference effects
in cognitive phenomena. Thus, the present Comment aims at formulating in precise
mathematical terms under what conditions the interference effects can arise and 
when they are prohibited. The clear understanding of these conditions will help 
avoiding incorrect interpretations and unnecessary misuse of the quantum 
formalism.

The idea that human psychological processes could be described by quantum theory
was advanced by Bohr (1933, 1958), who assumed that human cognition, involving
deliberations between several possible actions, could be modeled as a quantum
process. Deliberations between several complementary actions is analogous, Bohr 
argued, to interference between several quantum states. Since then, possible 
interference effects in human decision making have been discussed in a number 
of papers that suggested different models for resolving the so-called paradoxes 
in classical decision making and suggesting explanations of different cognitive 
phenomena. The history of this trend and the related models have been recently 
summarized in the book by Busemeyer and Bruza (2012).

As mentioned above, in the papers of the issue, interference is considered as 
a major quantum effect required for the explanation of a variety of cognitive 
phenomena. Even the order effects arising in different polls are also suggested 
as being connected with the manifestation of quantum features in decision making 
(Wang and Busemeyer, 2013), similarly to these effects in quantum phenomena.

The typical way of incorporating quantum effects into cognitive phenomena, which 
is employed by the majority of authors, is as follows. One first considers some 
particular cognitive empirical observations, then constructs a model for their 
explanation, postulating the occurrence of interference, and finally fits the 
model parameters to the same empirical observations.

This approach however is clearly unsatisfactory as a genuine scientific endeavor 
since the models: (a) lack generality, being mostly ad hoc, and (b) cannot claim 
predictive power as the same observations that motivated the models are used 
to qualify them. A real explanation of cognitive processes must be based on the 
following three necessary requirements:

(i) There should exist a general theory developed for arbitrary situations.
This theory has to be applicable to any cognitive phenomenon, instead of 
inventing particular models for each empirical case.

(ii) The theory has to clearly prescribe the conditions when quantum effects,
such as interference, appear and when they cannot arise. This prevents that 
interferences are postulated ad hoc.

(iii) The theory has to be able to make quantitative predictions without fitting
parameters. In other words, the process of validation must be developed, which
consists in making novel predictions to empirical situations that have not been
previously used to motivate the theory. Fitting a model to empirical observations 
can serve only as a first trivial step in the whole chain of model validation 
(Sornette et al., 2007, 2008).

In the next Sec. 2, we give a brief account of such a general theory. In Sec. 3,
to avoid confusion, we specify the terminology and notation employed in defining 
quantum events. Section 4 emphasizes the conditions when interference can arise 
and when it cannot. Particular models, which are suggested for interpreting 
cognitive phenomena, are to be tested against these conditions. Section 5 then 
considers the Gallup polls treated in the contribution by Wang and Busemeyer (2013) 
in the special issue, as an example illustrating our general classification.

\section{Decision making as a measurement procedure}

Formulating the theory of quantum measurements, von Neumann (1955) noticed that
a measurement procedure can be interpreted as a logical proposition satisfying 
the rules of quantum logic (Birkhoff and von Neumann, 1936). In quantum 
measurements, the observable measurable quantities are defined as the averages 
of self-adjoint operators from the algebra of local observables. More precisely, 
if an observable is represented by an operator $\hat{A}$ from the algebra of local 
observables, defined on a complex Hilbert space $\mathcal H$, then the observable 
quantity that can be measured is given by $\rm{Tr} \hat{\rho} \hat{A}$, where 
the trace is over $\mathcal H$ and $\hat{\rho}$ is a positive trace-one operator 
on $\mathcal H$, termed {\it statistical operator}, or {\it system state}, 
characterizing the considered system.

The rigorous definition of quantum probabilities can be done in line with 
the theory of quantum measurements. According to the Gleason (1957) theorem, 
a quantum probability measure has a unique extension to a positive linear 
functional defined for any bounded linear operator on a Hilbert space having 
a dimension larger than two, so that, if $A$ is a proposition, then the 
related quantum probability is uniquely given by $\rm{Tr} \hat{\rho} \hat{A}$.
Quantum probabilities, under different axiomatics, have been studied in a 
number of publications (Pitowsky, 1989). The achievable accuracy of quantum 
measurements has been analyzed by Dyakonov (2012). Psychological problems in 
the interpretation of quantum measurements have been discussed by Mermin (2012).

The most general and mathematically rigorous way of developing decision theory
with quantum probabilities is not by constructing some particular models, but by
following the general theory of quantum measurements. Such a quantum decision
theory has been recently advanced (Yukalov and Sornette, 2008) and developed in
a series of papers (Yukalov and Sornette, 2009a, 2009b, 2010a, 2010b, 2011). This
theory uniquely defines the conditions, when the interference effects can arise 
and when they do not occur. To explain this, we need to give a brief sketch of 
the approach, omitting mathematical details that can be found in the cited papers.

Consider a given set of prospects $\pi_j$, forming a complete lattice
$$
\cL = \{ \pi_j : \; j=1,2,\ldots, L \} \;  ,
$$
where the prospects are ordered by means of their probabilities, to be given
below. Each of the prospects is represented in a Hilbert space $\mathcal H$ as a
vector $|\pi_j \rangle$. The {\it prospect operators} on $\mathcal H$, defined as
\be
\label{1}
 \hat P(\pi_j) \equiv | \pi_j \rgl \lgl \pi_j | \;  ,
\ee
play the role of the operators of observables. Note that the prospect operators
are not required to be either orthogonal or to be projectors. A decision maker 
is characterized by a {\it decision-maker state} $\hat{\rho}$ on $\mathcal H$. 
The observable quantities, given by the averages
\be
\label{2}
 p(\pi_j) \equiv {\rm Tr} \hat\rho  \hat P(\pi_j) \;  ,
\ee
define the {\it prospect probabilities}. Note that the state $\hat{\rho}$ can be
either pure, if the decision maker is isolated, or mixed, when the decision maker
interacts with a surrounding society (Yukalov and Sornette, 2012).

Accomplishing the trace operation in definition (2), and separating the diagonal
and non-diagonal terms, results in the expression
\be
\label{3}
 p(\pi_j) = f(\pi_j) + q(\pi_j) \;  ,
\ee
where the first term corresponds to the diagonal representation, while the 
second term gives the off-diagonal contribution. The latter term is caused by 
quantum interference, so that we refer to it as the {\it interference factor} 
or {\it coherence factor}.

According to the generalized correspondence principle, first advanced by 
Bohr (1913) when analyzing atomic spectra, quantum theory must reduce to 
classical theory in the limit where quantum effects become negligible. In the 
more general version, the quantum-classical correspondence principle is 
understood as the requirement that the results of quantum measurements would 
be reducible to those of classical measurements when the quantum effects, 
such as interference, vanish. This reduction is called {\it decoherence} 
(Wheeler and Zurek, 1983; Zurek, 2003).

By the quantum-classical correspondence principle, the quantum probability, 
which is a measurable quantity, has to reduce to the classical probability, when 
the interference (coherence) factor tends to zero:
\be
\label{4}
p(\pi_j) \ra f(\pi_j) \; , \qquad q(\pi_j) \ra  0 \; .
\ee
Therefore, the diagonal term $f(\pi_j)$ corresponds to the classical probability.
The latter can be understood as a frequentist probability associated with the 
utility factor representing the objective utility of the prospect.

The interference term $q(\pi_j)$ is caused by the quantum nature of probability (2).
In decision theory, this term represents the subjective inclinations of a decision
maker caused by his/her emotions as well as possible behavioral biases, because of
which it can also be called the {\it attraction factor}.

It is clear that subjective feelings can vary in a wide range, being rather
different for different decision makers. Even for the same decision maker, under
the same conditions, with the same information, but at different times, subjective
feelings often change (Anderson et al., 1992; Anand et al., 2010). This, it seems,
would make it impossible to quantify the attraction factor. However, it has been 
shown (Yukalov and Sornette, 2009a, 2009b, 2010b, 2011) that the attraction factor,
arising in the definition of quantum probability (2), enjoys the following general
properties. It varies in the interval
\be
\label{5}
 -1 \leq q(\pi_j) \leq 1 \;  .
\ee
The following {\it alternation condition} holds:
\be
\label{6}
 \sum_{j=1}^L q(\pi_j) = 0 \;  .
\ee
And the average absolute value of the {\it aggregate} attraction factor can be 
estimated by the {\it quarter law}:
\be
\label{7}
  \frac{1}{L} \sum_{j=1}^L  | q(\pi_j) | = \frac{1}{4} \; ,
\ee
provided not all terms are zero.

As has been shown in our papers (Yukalov and Sornette, 2009a, 2009b, 2011), 
the quarter law holds true in a variety of experimental observations with a 
very good accuracy. However it is important to emphasize that this quarter law 
is not merely an empirical fact, but it can be theoretically derived as a 
{\it non-informative prior} (Yukalov and Sornette, 2009b, 2011). Because of the 
importance of the quarter law for quantitative prediction, we briefly sketch 
its derivation. Recall that the attraction factor $q$ is, generally, different 
for different decision makers, being in that sense a random quantity. Let 
$\vp (q)$ be a distribution of the attraction factor over the manifold of 
attraction factors corresponding to a large set of decision makers and/or of 
their mental states. According to the factor range (5), the distribution 
$\vp (q)$ is to be normalized over the domain $[-1,1]$ and has to satisfy the 
alternation condition (6), so that
$$
\int_{-1}^1 \vp(q) \; dq =  1 \; , \quad  \int_{-1}^1 \vp(q) q\; dq =  0 \; .
$$
Then the values
$$
q_+ \equiv  \int_0^1 \vp(q) q\; dq \; , \quad
q_- \equiv  \int_{-1}^0 \vp(q) q\; dq \; , 
$$
in view of the alternation condition, are related as $q_+ + q_- = 0$. The 
absence of any a priori information implies that the distribution 
$\varphi (q)$ is uniform. Then the normalization condition yields 
$\varphi (q) = 1/2$. As a result, from the above definition, we have $q_+ = 1/4$ 
and $q_- = - 1/4$. Thus, the non-informative prior for the absolute value of 
the aggregate attraction factor is $1/4$.  
  
These properties make it straightforward to give {\it quantitative predictions},
without any fitting parameters, for the behavior of decision makers. For 
instance, it is easy to give {\it quantitative} explanations for classical 
paradoxes in decision making, such as the disjunction effect or the conjunction 
fallacy (Yukalov and Sornette, 2009b, 2011).

\section{On terminology and notation for quantum events}

Defining quantum events or quantum measurements, one has to be accurate with
terminology and notation that may differ in the works of different authors.
To avoid misunderstanding, it is necessary to provide some precisions on our
terminology and notations. We shall be very brief, not going into mathematical 
details that can be found in the cited papers.

In quantum theory, an event $A$ is represented by a self-adjoint operator 
$\hat{A}$. For defining the probability of the event, one usually interprets
the latter operator as a projector. There are then two ways of introducing 
quantum probabilities. 

One way is to consider all operators, representing events, being defined on 
the same Hilbert space ${\cal H}$, with the system state $\hat \rho$. Then the 
probability of each event $A$ is $p(A) = {\rm Tr} \hat{\rho} \hat {A}$, with 
the trace over ${\cal H}$. For two disjoint events, say $A$ and $B$, their union 
$A \bigcup B$ is represented by the sum of the orthogonal operators 
$\hat A + \hat B$ acting on ${\cal H}$. The probability of this union 
$p(A \bigcup B)$ is the sum $p(A) + p(B)$.  
 
One calls the events disjoint, or mutually exclusive, or orthogonal, such that 
$A \bigcap B = 0$, when the related operators are mutually orthogonal projectors. 
In classical probability theory, the disjoint events are often termed as 
incompatible. The latter term, however, has a different meaning in quantum theory, 
where one calls incompatible, or incommensurable, such events whose related 
operators do not commute. And events are named compatible, or commensurable, 
when their operators commute. Therefore, to avoid confusion, we shall use the 
term {\it disjoint} for the orthogonal events, such that $A \bigcap B = 0$, while 
we shall call the events {\it compatible} when $[\hat A, \hat B] = 0$. Respectively, 
two events are incompatible, if the related operator commutator $[\hat A, \hat B]$ 
is not zero. 

In the procedure of quantum measurements of two incompatible observables, 
say $A$ and $B$, the order of their measurement is important, so that measuring 
$A \bigcap B$, generally, is not the same as measuring $B \bigcap A$. In order 
to avoid ambiguity, it is necessary to emphasize that, according to the standard 
notation accepted in quantum theory, the order of operations is understood as 
being accomplished from right to left. That is, when one writes the operations 
as ${\hat A} {\hat B}$, one assumes that $\hat B$ acts first, while $\hat A$ acts 
second. This order of actions is natural for quantum theory where actions are 
represented by operators acting on functions from a Hilbert space. Thus, the 
product of operators ${\hat A} {\hat B}$, acting on a function $\psi$, is uniquely 
defined as ${\hat A} \cdot {\hat B} \psi \equiv {\hat A} ({\hat B} \psi)$. 
We always use this standard definition.

There exists the well known problem in defining the joint probability of two 
incompatible events. In this respect, it has been shown (Niestegge, 2008) that 
the quantum joint probability $p(A \bigcap B)$ can be correctly defined either 
for compatible events, for which $\hat A$ and $\hat B$ commute, or for incompatible 
events by assigning the operators to the Jordan algebra, where the operator 
product ${\hat A} \cdot {\hat B}$ is given by the Jordan symmetric form 
$({\hat A}\cdot {\hat B}+{\hat B}\cdot {\hat A})/2$. However, such a symmetric 
definition of the joint probability possesses the properties of the classical
probability yielding no interference terms. 

When an event $A$ occurs after an event $B$, then, instead of the joint 
probability, one considers the transition probability defined as the product 
of the L\"{u}ders projection-rule probability of successive measurements 
$p_L(A|B)$ and the probability $p(B)$. This transition probability was first 
introduced by Wigner (1932), so that one can denote it as 
$p_W(A|B) = p_L(A|B) p(B)$. It is this Wigner transition probability that is 
used by Wang and Busemeyer (2013). The L\"{u}ders probability is often 
interpreted as an extension to the quantum region of the classical conditional 
probability. However, when the events $A$ and $B$ are represented by one-dimensional 
orthogonal projectors, the L\"{u}ders form $p_L(A|B)$ is symmetric, such that 
$p_L(A|B) = p_L(B|A)$. Contrary to this, the classical conditional probabilities 
are not symmetric. Therefore, the L\"{u}ders form cannot be treated as an extension 
of the classical conditional probability. As a consequence, the Wigner transition 
probability cannot be accepted as an extension of the joint probability.  

There is another way of introducing quantum probabilities for several events,
which is free of the problem of defining the joint probabilities. In this 
approach, each event $A$ is represented by a proposition operator ${\hat A}$ 
acting on a Hilbert space ${\cal H}_A$. Respectively, another event $B$ is 
represented by an operator ${\hat B}$ acting on a Hilbert space ${\cal H}_B$. 
The system state is defined by a statistical operator $\hat \rho_{AB}$ acting on 
the tensor-product space ${\cal H}_{AB} \equiv {\cal H}_A \bigotimes {\cal H}_B$. 
The probability of a separate event 
$p(A) \equiv {\rm Tr} {\hat \rho_{AB}} {\hat A}$, with the trace over the tensor 
product ${\cal H}_{AB}$, reduces to the previous definition above. 

The sequence of two events $A$ and $B$ is called a {\it composite event} and 
is denoted by $A \bigotimes B$ or simply as $AB$, and is represented by the 
tensor-product operator ${\hat A} \bigotimes {\hat B}$ acting on ${\cal H}_{AB}$. 
By this definition, the joint probability of these two events is 
$p(AB)={\rm Tr}{\hat \rho_{AB}}{\hat A} \bigotimes {\hat B}$, with the trace 
over the tensor-product space ${\cal H}_{AB}$. It is this definition of composite 
events that we have used from the very beginning in our approach to quantum decision 
theory (Yukalov and Sornette, 2008, 2009a, 2009b, 2010a, 2010b, 2011). Further 
mathematical information on the properties of composite events can be found in
the literature (Holevo, 1973; Wilce, 1992; Niestegge, 2004; Harding, 2009).                
 
In the majority of cases studied in experiments performed in cognitive sciences, 
one considers the situation where the prospect lattice is formed by just two 
prospects, $\pi_1 = A_1 B$ and $\pi_2 = A_2 B$, where $B \equiv \{B_1, B_2\}$ is 
represented as a set of two possible actions (or events), $B_1$ and $B_2$.
By the general theory, for this case, the prospect states are given by 
$|\pi_j \rangle = |A_j \rangle \bigotimes |B \rangle$. The prospect probabilities
are
\be
\label{8}
p(A_jB) =  p(A_jB_1) + p(A_jB_2) + q(A_jB) \;  ,
\ee
where $j = 1, 2$. The first two terms, whose sum is 
$f(A_jB) = p(A_jB_1) + p(A_jB_2)$, are the classical joint probabilities that 
are uniquely defined through the conditional probabilities:
\be
\label{9}
 p(A_iB_j) =  p(A_i|B_j)  p(B_j)  .
\ee
Hence, the interference term (attraction factor) is
\be
\label{10}
 q(A_jB) = p(A_jB) - p(A_jB_1) -  p(A_jB_2)\;  .
\ee
The alternation condition (6) now reads as $q(A_1B) = - q(A_2B)$, and the 
quarter law (7) becomes $| q(A_1B) | + | q(A_2B) | = 1/2$.

Interference effects exist only when the interference factor (10) is not zero.
Only then, there is the necessity of invoking quantum notions for cognitive
phenomena. When this factor is zero, there is no interference effects and all
cognitive phenomena can be described with a classical formalism.

\section{Interference appears in decisions under uncertainty}

As has been mentioned, postulating the existence of interference effects may 
lead to incorrect conclusions, since interference does not necessarily occur 
for any prospect. Let us now clarify this and formulate the conditions under 
which interference cannot arise and those when it can. These conditions follow 
from the general theory delineated above. The mathematical details can be 
found in our earlier publications 
(Yukalov and Sornette, 2008, 2009a, 2009b, 2010b, 2011).

Prospects are composed of elements that, depending on applications, may be called
events (in probability theory), propositions (in logic), measurements (in natural
sciences), or decisions and actions (in decision theory). Keeping this in mind, we
use the term ``event".

\vskip 2mm

(i) {\it Simple events}.

\vskip 2mm

A simple event $A_i$ is such that it cannot be represented as a union or 
conjunction of several events. There is no interference for a simple event:
$$
 q(A_i) = 0 \;  .
$$

\vskip 2mm

(ii) {\it Unions of mutually disjoint simple events}.

\vskip 2mm

Let $\bigcup_i A_i$ be a union of mutually disjoint simple events $A_i$ all 
represented by orthogonal projectors on the same space $\cal H$. There is no 
interference for this union:
$$
 q \left (\bigcup_i A_i \right ) = 0 \;  .
$$

\vskip 2mm

(iii) {\it Factorized composite events}.

\vskip 2mm

An event is composite if it can be represented as a conjunction of several 
events as defined above. A composite event is factorized if it is a conjunction 
$\bigotimes_i A_i$ of simple events $A_i$, whose representing operators 
${\hat A}_i$ are defined each on its own space ${\cal H}_i$. There is no 
interference for these factorized composite events:
$$
q \left (\bigotimes_i A_i \right ) = 0 \;   .
$$

\vskip 2mm

(iv) {\it Entangled composite events}.

\vskip 2mm

A composite event is entangled if it cannot be represented as a factorized event.
An example of an entangled event is $A_i B$, where $B = \{B_1, B_2\}$, so that the
prospect state is $|A_i \rangle \bigotimes |B \rangle$. Interference can occur 
only for entangled events. For illustration, let us recall the double-slit 
experiment in physics. A particle is emitted in the direction of a screen having 
two slits. From another side of the screen, there are detectors registering the 
arrival of the particle. Let the registration of the particle by an $i$-detector 
be denoted as $A_i$ and the passage of the particle through one of the slits be 
denoted as $B_1$ or $B_2$, respectively. When the passage of the particle through 
a slit $B_j$ is certain, then the event $A_i B_j$ is factorized and displays no 
interference, with the event probability given by $p(A_i B_j)$ and the vanishing 
factor $q(A_i B_j)$.

However, when it is not known through which of the slits the particle passes, 
then the event $A_i B$, is entangled and demonstrates interference, that is, a 
nonzero factor $q(A_i B)$, giving rise to the remarkable quantum effect of 
interference fringes, or periodic modulations of the probability of finding the 
particle along the dimension of the screen.

\section{Is there anything quantum in Gallup polls?}

As an application of the above analysis, let us consider the Gallup polls treated
in the contribution by Wang and Busemeyer (2013). In accordance with the
considerations we have presented above, this contribution is the most interesting, 
since it attempts to give an explanation for the necessity of involving quantum 
notions in order to interpret empirical tests, such as Gallup polls, without 
introducing fitting parameters.

The Gallup polls, demonstrating order effects, have been described by Moore
(2002). In the polls, conducted in September 6, 1997, the respondents were asked
to answer the questions of the following type: "Do you generally think Clinton
(Gore) is honest and trustworthy?" The questions were asked in two separate
contexts: non-comparative and comparative. The non-comparative context for a 
question occurs when that question is asked first, without any mention of the 
other item. The comparative context for a question occurs when that question is 
asked after the other question in the pair, so that the second question may be 
influenced by the response to the first question. The Gallup polls, as described 
by Moore, have demonstrated noticeable order effects. Thus, when respondents were 
asked about Clinton first, $50\%$ said he was honest and trustworthy. When the 
other group of the sample was asked about Gore first, $68 \%$ said he was honest 
and trustworthy. However, in the comparative context, when the question about 
Clinton was asked second, $57 \%$ said he was honest. And when the question about 
Gore was asked second, $60 \%$ said he was honest. This demonstrated the 
questions-order effect, or more precisely, the difference between the comparative 
and non-comparative contexts.

At this point, it is worth noting that these numbers, to our mind, should not 
be treated as having absolute value. They are actually to a large extent rather 
random. For instance, the responses to the same questions about Clinton and Gore, 
given at very close times, fluctuated around $50\%$ for both of them, since, 
generally, people did not see much difference between these two democrats 
furthermore associated in the US Administration as president and vice-president
respectively (Gallup poll, 1997). For example, in the Gallup poll of October 3, 
1997, Clinton was considered as honest by $56\%$ of respondents and Gore, 
by $47\%$, which shows a relation opposite to the previous September poll, when 
respondents classified Gore as being more honest than Clinton. And in the poll 
of October 30, 1997, Clinton was considered honest by $62\%$ of respondents, 
while Gore, by $53\%$, again with a relation opposite to the September poll. 
This shows that these polls exhibit very strong fluctuations with time. While, 
at each {\it fixed} poll, the order effects were found to be statistically 
significant (Moore, 2002), this does not contradict the fact that a poll as a 
whole can be random, displaying different data at close times, when the external 
conditions are practically the same.   

The typical structure of the polls is as follows. Respondents first answer 
a question $B$, choosing either $B_1$ or $B_2$, and after this, answer a 
question $A$, choosing either $A_1$ or $A_2$. Suppose a population of $N$
respondents is interrogated. Answering the question $B$, a part $N(B_i)$ 
answers $B_i$, with $N(B_1) + N(B_2) = N$. The corresponding fractions give 
the unconditional frequentist probabilities $p(B_j) = N(B_j)/N$. Then each 
part $N(B_j)$ of the population is questioned on $A$, which separates the 
population into the subpopulations $N(A_i|B_j)$ of those who answer $A_i$ 
under the condition that before they have answered $B_j$. Clearly, 
$N(A_1|B_j) + N(A_2|B_j) = N(B_j)$. The related fractions define the classical 
{\it conditional} probabilities $p(A_i|B_j) = N(A_i|B_j)/N(B_j)$, satisfying 
the standard normalization for classical conditional probabilities
$p(A_1|B_j) + p(A_2|B_j) = 1$. Classical joint probabilities can be defined 
as $p(A_iB_j) = p(A_i|B_j) p(B_j)$. In the same way, the other group of 
respondents is interrogated, first on $A$ and then on $B$, giving the classical probabilities $p(B_jA_i)$. 
     
The classical joint probabilities are symmetric, so that
$p(A_iB_j) = p(B_jA_i)$ should hold. But the Gallup polls have shown that this 
is not always the case. To explain the order effects, Wang and Busemeyer (2013) 
suggested to treat $p(A_iB_j)$ as the Wigner transition probability
$p_W(A_i|B_j) = p_L(A_i|B_j)p(B_j)$. Their main result, that they call the 
"q-test", is the demonstration that the so interpreted probabilities satisfy 
the equality
\be
\label{11}
p(A_1B_2) + p(A_2B_1) = p(B_1A_2) +  p(B_2A_1) \;  ,
\ee
which is in good agreement with the Gallup polls. 

However, to satisfy this equality, there is no need to resort to quantum 
notions. This is because Eq. (11) is also valid for classical probabilities
due to their symmetry. Therefore, the validity of this test, as such, does not 
prove yet the quantum origin of the phenomenon. What is specific for the Gallup 
polls is not this equation, but the asymmetry of the related frequentist 
probabilities. That is, the proof would be if it could be possible to calculate
somehow quantum probabilities $p_{quantum}(A_iB_j)$ and to show that they 
coincide with the empirical frequencies $p(A_iB_j)$.   
 
The events $A_i$ or $B_j$, from the mathematical point of view defined above, 
are not entangled. Consequently, the corresponding probabilities $p(A_i)$ 
and $p(B_j)$, according to our classification, involve no interference. Thus, we
are not convinced of the need to invoke interference effects to explain Gallup polls, 
due to a lack of genuine uncertainty in the related decision making. While
Wang and Busemeyer (2013) mention that the questions posed in the polls
can contain some kind of uncertainty, the general theory that
we have outlined above specifies rigorously the type of uncertainty that ensures
the existence of interference: it is the mathematically defined uncertainty related 
to the occurrence of {\it entangled composite events}. In the absence of such events, 
there can be no interference terms. The analysis of the order effects by Wang and 
Busemeyer (2013) is certainly of interest. However, it remains unclear why 
the Gallup polls would require quantum interpretation. 

\vskip 2mm

In conclusion, the rigorous approach, based on the theory of quantum measurements, 
allows us to find the conditions when quantum effects, such as interference, can 
appear in cognitive sciences. Possible occurrence of quantum effects in cognitive 
empirical tests should be checked against these conditions. Generally, quantum 
effects can arise only when one makes decisions under uncertainty, in the presence 
of entangled composite events.

\newpage

{\bf References}

\vskip 5mm

{\parindent=0pt
Anand, K., Gai, P., and Marsili, M. (2010).
The rise and fall of trust networks.
{\it Lecture Notes in Economics and Mathematical Systems, 645}, 77--88.

\vskip 2mm
Anderson, S.P., De Palma, A., and Thisse, J.F. (1992).
{\it Discrete Choice Theory of Product Differentiation}.
Cambridge: Massachusetts Institute of Technology.

\vskip 2mm
Birkhoff, G., and von Neumann, J. (1936).
The logic of quantum mechanics.
{\it Annals of Mathematics, 37}, 823--843.

\vskip 2mm
Bohr, N. (1913).
On the constitution of atoms and molecules.
{\it Philosophical Magazine, 26}, 1--25, 476--502, 857--875.

\vskip 2mm
Bohr, N. (1933).
Light and life. {\it Nature, 131}, 421--423, 457--459.

\vskip 2mm
Bohr, N. (1958).
{\it Atomic Physics and Human Knowledge}. New York: Wiley.

\vskip 2mm
Busemeyer, J.R., and Bruza, P. (2012).
{\it Quantum Models of Cognition and Decision}. Cambridge: Cambridge University.

\vskip 2mm
Dyakonov, M.I. (2012).
State of the art and prospects for quantum computing.
arXiv:1212.3562.

\vskip 2mm
Gallup poll (1997).
edition.cnn.com/ALLPOLITICS/1997/10/30/poll/

\vskip 2mm
Gleason, A.M. (1957).
Measures on the closed subspaces of a Hilbert space.
{\it Journal of Mathematics and Mechanics, 6}, 885--893.

\vskip 2mm
Harding, J. (2009).
A link between quantum logic and categorical quantum mechanics.
{\it International Journal of Theoretical Physics, 48}, 769--802. 

\vskip 2mm
Holevo, A.S. (1973).
Statistical decision theory for quantum systems.
{\it Journal of Multivariate Analysis, 3}, 337--394.

\vskip 2mm
Mermin, N.D. (2012).
Quantum mechanics: fixing the shifty split.
{\it Physics Today, 65}, 8--10.

\vskip 2mm
Moore, D.W. (2002).
Measuring new types of question-order effects.
{\it Public Opinion Quarterly, 66}, 80--91.

\vskip 2mm
Niestegge, G. (2004).
Composite systems and the role of the complex numbers in quantum mechanics.  
{\it Journal of Mathematical Physics, 45}, 4714--4725.

\vskip 2mm
Niestegge, G. (2008)
An approach to quantum mechanics via conditional probabilities.
{\it Foundations of Physics, 38}, 241--256.  

\vskip 2mm
Pitowsky, I. (1989).
{\it Quantum Probability--Quantum Logic}. Berlin: Springer.

\vskip 2mm
Sornette, D., Davis, A.B., Ide, K., Vixie, K.M.R., Pisarenko, V., and Kamm, J.R. (2007).
Algorithm for model validation: theory and applications.
{\it Proceedings of National Academy of Sciences of USA, 104}, 6562--6567.

\vskip 2mm
Sornette, D., Davis, A.B., Kamm, J.R., and Ide, K. (2008).
A general strategy for physics-based model validation illustrated with
earthquake phenomenology, atmospheric radiative transfer, and computational
fluid dynamics.
{\it Lecture Notes in Computational Science and Engineering, 62}, 19--73.

\vskip 2mm
von Neumann, J. (1955).
{\it Mathematical Foundations of Quantum Mechanics}.
Princeton: Princeton University.

\vskip 2mm
Wang, Z., and Busemeyer, J.R. (2013).
A quantum question order model supported by empirical tests of an a priori 
and precise prediction. 
{\it Topics in Cognitive Science} (in press). 

\vskip 2mm
Wang, Z., Busemeyer, J.R., Atmanspacher, H., and Photos, E. (2013). 
The potential of using quantum theory to build models of cognition.
{\it Topics in Cognitive Science} (in press). 

\vskip 2mm
Wheeler, J.A., and Zurek, W.H. (1983).
{\it Quantum Theory and Measurement}. Princeton: Princeton University.

\vskip 2mm
Wigner, E. (1932).
On the quantum correction for thermodynamic equilibrium.
{\it Physical Review, 40}, 749--759. 

\vskip 2mm
Wilce, A. (1992).
Tensor products in generalized measure theory.
{\it International Journal of Theoretical Physics, 31}, 1915--1928.

\vskip 2mm
Yukalov, V.I., and Sornette, D. (2008).
Quantum decision theory as quantum theory of measurement.
{\it Physics Letters A, 372}, 6867--6871.

\vskip 2mm
Yukalov, V.I., and Sornette, D. (2009a).
Physics of risk and uncertainty in quantum decision making.
{\it European Physical Journal B, 71}, 533--548.

\vskip 2mm
Yukalov, V.I., and Sornette, D. (2009b).
Processing information in quantum decision theory.
{\it Entropy, 11}, 1073--1120.

\vskip 2mm
Yukalov, V.I., and Sornette, D. (2010a).
Entanglement production in quantum decision making.
{\it Physics of Atomic Nuclei, 73}, 559--562.

\vskip 2mm
Yukalov, V.I., and Sornette, D. (2010b).
Mathematical structure of quantum decision theory.
{\it Advances in Complex Systems, 13}, 659--698.

\vskip 2mm
Yukalov, V.I., and Sornette, D. (2011).
Decision theory with prospect interference and entanglement.
{\it Theory and Decision, 70}, 283--328.

\vskip 2mm
Yukalov, V.I., and Sornette, D. (2012).
Quantum decision making by social agents.
{\it Swiss Finance Institute Research Paper No. 12--10},
http://ssrn.com/abstract=2018270.

\vskip 2mm
Zurek, W.H. (2003).
Decoherence, selection, and the quantum origins of the classical.
{\it Reviews of Modern Physics, 75}, 715--775.

}

\end{document}